\documentclass[twocolumn,showpacs,preprintnumbers,amsmath,amssymb]{revtex4}
\usepackage{graphicx}
\usepackage{textcomp}

\begin{document}
\draft
\title{Dynamic localization and transport in complex crystals}
\normalsize

\author{S. Longhi}
\address{Dipartimento di Fisica and Istituto di Fotonica e
Nanotecnologie del CNR, Politecnico di Milano, Piazza L. da Vinci 32, I-20133 Milano, Italy}


%
\bigskip
\begin{abstract}
\noindent The behavior of a Bloch particle in a complex crystal with
$\mathcal{PT}$ symmetry subjected to a sinusoidal ac force is
theoretically investigated. For unbroken $\mathcal{PT}$ symmetry and
in the single-band approximation, it is shown that time reversal
symmetry of the ac force preserves the reality of the quasienergy
spectrum. Like in ordinary crystals, exact band collapse,
corresponding to dynamic localization, is attained for a sinusoidal
band shape. The wave packet dynamics turns out to be deeply modified
at the $\mathcal{PT}$ symmetry breaking point, where band merging
occurs and Bragg scattering in the crystal becomes highly
non-reciprocal.
\end{abstract}



\maketitle
\section{Introduction}
 The dynamics of matter or classical waves in
periodic potentials subjected to external dc or ac forces is
strongly influenced by Bragg scattering, which is responsible for
such important effects as Bloch oscillations (BOs) and dynamic
localization (DL). Dynamic localization was originally proposed by
Dunlap and Kenkre \cite{Dunlap86} as a suppression of the broadening
of a charged-particle wave packet as it moves along a tight-binding
lattice driven by a sinusoidal ac electric field. DL was then
explained in terms of quasienergy band collapse \cite{Holthaus92},
and the general conditions of DL beyond the nearest-neighboring
tight-binding
 (NNTB) approximation used by Dunlap and Kenkre
were subsequently investigated in Ref.\cite{Dignam02}. The interest
on DL has been recently renewed since the first experimental
observations of DL have been reported for matter waves trapped in
dynamical optical lattices \cite{Madison98,Ligner07,Eckardt09}, and
for light waves in curved waveguide arrays
\cite{Longhi06,Longhi06b,Iyer07,Dreisow08DL,Szameit09}. Matter or
light waves may also interact with complex potentials. Complex
crystals for matter waves emerge, for instance, in the near resonant
interaction of light with open two-level systems \cite{Keller97},
whereas in optics complex crystals are realized by waveguide arrays
that include gain and/or loss regions \cite{Makris08}. As compared
to ordinary crystals, complex crystals exhibit some unique
properties, such as violation of the Friedel's law of Bragg
scattering and nonreciprocal diffraction
\cite{Makris08,Keller97,Berry98}. A special class of complex
crystals is provided by complex potentials possessing parity-time
($\mathcal{PT}$) symmetry \cite{Makris08,Bender1,Bender2}. An
important property of $\mathcal{PT}$ crystals is to admit of an
entirely real-valued energy spectrum below a phase transition
(symmetry-breaking) point, in spite of the non-Hermiticity of the
underlying Hamiltonian \cite{Bender1}. A recent study on BOs in
$\mathcal{PT}$ crystals \cite{Longhi09} has shown that the common
wisdom of coherent quantum transport in a crystal is greatly
modified
when dealing with a complex crystal.\\
 It is the aim of this work
to investigate the coherent motion of wave packets in a complex
$\mathcal{PT}$ crystal driven by an ac-like force. In particular, it
is shown that in the unbroken $\mathcal{PT}$ symmetry region time
reversal symmetry of the ac-like force preserves the reality of the
quasienergy spectrum, and that a full band collapse, corresponding
to dynamic localization, occurs within the single-band and NNTB
approximations like in ordinary crystals. However, the transport
properties of the lattice are deeply modified at the $\mathcal{PT}$
symmetry breaking, where Bragg scattering in the crystal becomes
highly non-reciprocal.  In the following analysis, we will consider
specifically wave packets in a photonic lattice system
\cite{Longhi06,Szameit09,Makris08,Longhi09}, however the results can
be applied to other lattice realizations, such as to matter wave
tunneling in dynamic complex optical lattices.

\section{Dynamic localization in complex crystals with unbroken $\mathcal{PT}$ symmetry}
 In optics, the
coherent motion of charged quantum particles in periodic potentials
driven by an ac electric field can be mimicked by the propagation of
light waves in a periodically-curved photonic lattice
\cite{Longhi06,Longhi09b}. In the scalar and paraxial
approximations, light propagation at wavelength $\lambda$ in the
lattice is described by the Schr\"{o}dinger-type wave equation
\begin{equation}
i \lambdabar \partial_z \psi= -\frac{\lambdabar^2}{2n_s}
\partial^2_x \psi+V(x)\psi -Fx \psi \equiv (\mathcal{H}_0-Fx)\psi.
\end{equation}
where $\lambdabar=\lambda/(2 \pi)$ is the reduced wavelength, $n_s$
is the substrate refractive index, $V(x)=n_s-n(x)$ is the potential,
$n(x)=n(x+a)$ is the refractive index profile of the lattice
(spatial period $a$), and $F=F(z)$ is a fictitious refractive index
gradient proportional to the local waveguide axis curvature which
mimics the action of a driving force \cite{Longhi09b}. In
particular, a sinusoidal ac-like force is simply mimicked by a
sinusoidal bending profile of the waveguides \cite{Longhi06}. In a
complex lattice, the refractive index is complex, and the
$\mathcal{PT}$ symmetry requirement
 $V(-x)=V^*(x)$ corresponds to suitable combinations of optical gain and loss regions in the
lattice as discussed in \cite{Makris08}. The real and imaginary
parts of the potential are denoted by $V_R(x)$ and $\alpha V_I(x)$,
respectively, where $\alpha \geq 0$ is a dimensionless parameter
that measures the anti-Hermitian strength of $\mathcal{H}_0$. The
spectrum of $\mathcal{H}_0$ turns out to be real-valued for $\alpha
< \alpha_{c}$, where $\alpha_c \geq 0$ corresponds to the transition
from unbroken to broken $\mathcal{PT}$ symmetry. Numerical studies
generally show that for $\alpha < \alpha_c$ the spectrum is composed
by bands separated by gaps like in an ordinary crystal, whereas for
$\alpha \geq \alpha_c$ band merging is observed with the appearance
of pairs of complex-conjugate eigenvalues \cite{Makris08,Bender2}.
For instance, for the potential defined by
\begin{equation}
V_R(x)=V_0 \cos(2 \pi x/a) \; , \; \; V_I(x)=V_0 \sin(2 \pi x/a),
\end{equation}
 one has $\alpha_c=1$ \cite{Makris08}.\\
In this section we consider the unbroken symmetry phase, i.e. the
case $\alpha<\alpha_c$. As recently shown in Ref.\cite{Longhi09},
for $\alpha<\alpha_c$ the motion of a Bloch particle in presence of
an external dc force $F$ can be described following the same lines
as in ordinary crystals by expanding the field $\psi(x,z)$ as a
superposition of Bloch-Floquet eigenfunctions
$\phi_n(x,\kappa)=u_{n}(x,\kappa) \exp(i \kappa x)$ of
$\mathcal{H}_0$, where the wave number $\kappa$ varies in the first
Brillouin zone, i.e. $-k_B/2 \leq \kappa < k_B/2$, $k_B=2 \pi/a$ is
the Bragg wave number, $n$ is the band index, and $u_n(x,\kappa)$ is
the periodic part of the Bloch-Floquet eigenfunction. After setting
$\psi(x,z)=\sum_n \int d \kappa c_n(\kappa,z) \phi_n(\kappa,z)$ and
assuming normalized eigenfunctions such that $ \int dx
\phi_{n'}^*(-x,-\kappa') \phi_n(x,\kappa)=\mathcal{D}_n
\delta_{n,n'} \delta(\kappa-\kappa')$ with $\mathcal{D}_n= \pm 1$,
the evolution equations for the spectral coefficients
$c_n(\kappa,z)$ read \cite{Longhi09}
\begin{equation}
i \lambdabar \left( \partial_z+\frac{F}{\lambdabar}
\partial_{\kappa} \right)c_n = E_n(\kappa)c_n-F \mathcal{D}_n \sum_l X_{n,l}(\kappa)
c_l
\end{equation}
where $E_n(\kappa)$ is the energy of $\phi_n(x,\kappa)$ [with
$E_n(-\kappa)=E_n(\kappa)$] and $X_{n,l}(\kappa) \equiv (2 \pi i /a)
\int_{0}^{a}dx u_{n}^{*}(-x,-\kappa)
\partial_{\kappa} u_l(x,\kappa)$. The off-diagonal elements
$X_{n,l}$ ($ n \neq l$) in Eq.(3) are responsible for interband
transitions, i.e. Zener tunneling (ZT). If bands $n$ and $l$ are
separated by a large gap and the ac force $F(z)$ is small enough
such that $|F X_{n,l}(\kappa)| \ll |E_{n}(\kappa)-E_l(\kappa)|$ in
the entire Brillouin zone, ZT is negligible as in ordinary lattices
and one can make the single-band approximation by setting $X_{n,l}
\simeq 0$ for $n \neq l$ in Eq.(3). In the single-band approximation
one thus obtains
\begin{equation}
i \lambdabar \left( \partial_z+\frac{F}{\lambdabar}
\partial_{\kappa} \right)c(z,\kappa)= \left[ E (\kappa) -i F(z) \Phi(\kappa)
\right] c(\kappa,z)
\end{equation}
where we omitted, for the sake of simplicity, the band index $n$ and
set $i \Phi(\kappa) \equiv \mathcal{D}_nX_{n,n}(\kappa)$. Because of
the symmetry of $V(x)$, $\mathrm{Re}(u_n(k,x))$ and
$\mathrm{Im}(u_n(k,x))$ have well defined and opposite parity under
the inversion $x \rightarrow -x$; this implies that $\Phi(\kappa)$
is a real-valued function of $\kappa$, vanishing for a real
potential (i.e. for $\alpha=0$). As previously shown in
\cite{Longhi09}, when a dc force $F$ is applied to the crystal, from
Eq.(4) it follows that the energy spectrum is described by a
complex-valued Wannier-Stark ladder. The non-reality of the energy
spectrum comes from the extra-term $\Phi(\kappa)$ in Eq.(4) and is
physically due to the fact that the external dc force $F$ breaks the
$\mathcal{PT}$ symmetry of the full Hamiltonian
$\mathcal{H}=\mathcal{H}_0-Fx$. For an ac-like force with period
$\Lambda= 2 \pi/ \omega$, because of the $z$-periodicity of the
Hamiltonian its energy spectrum is replaced by a quasi-energy
spectrum. Moreover, in the single-band approximation DL corresponds
to a complete collapse of the quasi-energy band like in an ordinary
crystal \cite{Holthaus92}. According to Floquet's theorem of
periodic systems, the quasi-energy $\mathcal{E}(\kappa)$ for the
$n$-th lattice band  can be readily calculated by looking for a
solution to Eq.(4) of the form $c(z,\kappa)=a(z,\kappa) \exp[- i
\mathcal{E}(\kappa)z / \lambdabar]$ with
$a(z+\Lambda,\kappa)=a(z,\kappa)$. One obtains
\begin{equation}
\mathcal{E}(\kappa)=\frac{1}{\Lambda} \int_0^{\Lambda} dz \left[ E
(\kappa')-i F(z) \Phi(\kappa') \right]
\end{equation}
where we have set $\kappa' \equiv \kappa-k(\Lambda)+k(z)$ and
$k(z)=(1 / \lambdabar) \int_0^z d \xi F(\xi)$. Let us assume that
the ac forcing $F(z)$ is an odd function with respect to some point
$z_0$, i.e. that $F(z-z_0)=-F(z_0-z)$ for some $z_0$ in the
oscillation cycle. This condition is satisfied, for instance, for
the important case of an harmonic (e.g. sinusoidal or cosinusoidal)
ac driving force, originally considered by Dunlap and Kenkre
\cite{Dunlap86} and that will be assumed in the following. Owing to
this additional temporal symmetry on the driving force, which is
absent for the BO problem \cite{Longhi09}, a real-valued
quasi-energy spectrum for the non-Hermitian time-periodic
Hamiltonian $\mathcal{H}_0-F(z)x$ is obtained. In fact, in this case
the imaginary term on the right hand side of Eq.(5) vanishes after
integration because $F(z)$ and $\Phi(\kappa')$ have opposite parity
for the inversion $(z-z_0) \rightarrow -(z-z_0)$. The quasi-energy
spectrum is thus real-valued and its expression takes the usual form
as in a conventional crystal. DL corresponds to a collapse of the
quasienergy band $\mathcal{E}(\kappa)$, i.e. to $ d
\mathcal{E}(\kappa) / d \kappa=0$. For most driving fields like a
sinusoidal field, DL can be attained exactly solely in the NNTB
approximation, i.e. when the band shape $E(\kappa)$ is sinusoidal
\cite{Dignam02}, $E(\kappa)=E_0-\Delta \cos(\kappa a)$. In this
case, assuming for the sake of definiteness a sinusoidal ac-like
force $F(z)=F_0 \cos(\omega z)$, the explicit form of the
quasi-energy
 reads \cite{Holthaus92}
\begin{equation}
\mathcal{E}(\kappa)=E_0-\Delta J_0 \left( \frac{F_0 a}{\lambdabar
\omega} \right) \cos(\kappa a).
\end{equation}
Band collapse, leading to DL, is thus attained when $J_0 ( F_0 a /
\lambdabar
\omega)=0$  \cite{Dunlap86,Holthaus92,Longhi06}.\\
\begin{figure}[htbp]
  \includegraphics[width=85mm]{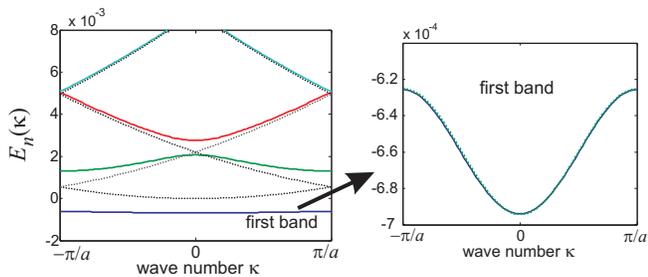}\\
   \caption{(color online) Band diagram of the complex potential defined by Eq.(2) for parameter values
   $\lambda=633$ nm, $a=8 \; \mu$m,  $n_s=1.42$, $V_0=0.002$, and $\alpha=0.3$. The dotted curve shows, for comparison, the
   parabolic dispersion relation $E(\kappa)=(\lambdabar \kappa)^2/(2 n_s)$, folded inside the first Brillouin
   zone, corresponding to the critical case $\alpha=\alpha_c=1$. The
   picture on the right hand side is an enlargement of the lowest
   band (solid curve), fitted by a sinusoidal curve
   (dotted curve), almost overlapped with the solid one.}
\end{figure}
\begin{figure}[htbp]
  \includegraphics[width=85mm]{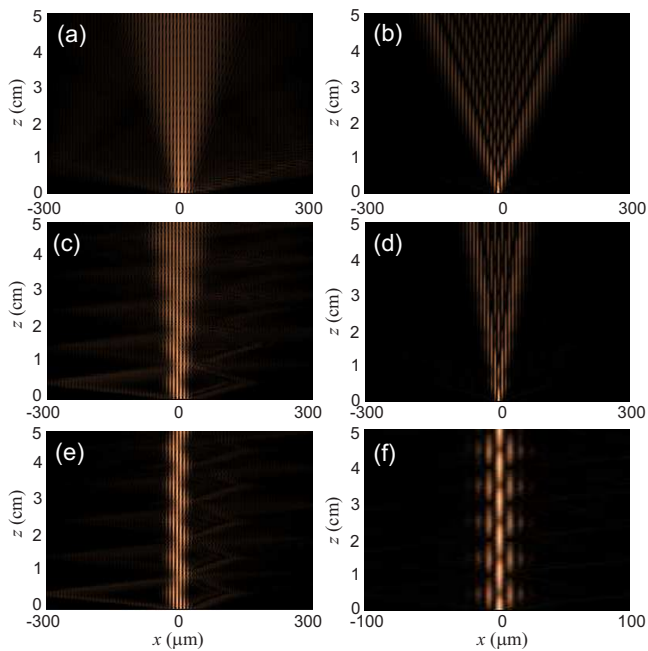}\\
   \caption{(color online) Wave packet broadening (discrete diffraction) and dynamic localization in the lattice of Fig.1
   for broad beam excitation at normal incidence (left panels) and for single
   site excitation at normal incidence (right panels). In (a) and
   (b), the external force is absent ($F_0=0$). In (c), (d) a
   sinusoidal force with period $\Lambda=1$ cm and amplitude $F_0=13.32 \; \mu {\rm
   m}^{-1}$, corresponding to $\Gamma=1.684$, is applied. In (e) and (f), the forcing amplitude is increased to $F_0=19.03 \; \mu {\rm
   m}^{-1}$, corresponding to the DL condition $\Gamma=2.405$.}
\end{figure}
To check the correctness of the analysis, we investigated DL for the
complex crystal $V(x)=V_0[\cos(2 \pi x/a)+i \alpha \sin(2 \pi x/a)]$
in the unbroken $\mathcal{PT}$ symmetry phase ($\alpha<1$) by a
direct numerical analysis of Eq.(1) using a pseudo-spectral
split-step method with absorbing boundary conditions. Figure 1 shows
a typical example of band diagram below the phase transition point
$(\alpha=0.3$), numerically computed by a spectral analysis of the
Hamiltonian $\mathcal{H}_0$. Note that the lowest band of the array
is with excellent accuracy approximated by a sinusoidal curve and
turns out to be separated by a large gap from the second band. DL is
thus expected to occur provided that the lattice is excited in its
lowest order band. Figures 2(a) and (b) show a typical spreading
(discrete diffraction) of a Gaussian wave packet
$\psi(x,0)=\exp(-x^2/w^2)$ in the absence of the external force, for
either a broad Gaussian beam that excites at normal incidence a few
lattice sites at $z=0$ [$2w=5a$, Fig.2(a)], and a narrow Gaussian
beam that excites at normal incidence a single well of the lattice
[$2w=2a/3$, Fig.2(b)]. In both cases, the lowest band of the array
is mainly excited, as discussed e.g. in \cite{Longhi06b}. After
application of the sinusoidal force [Figs.2(c-f)], suppression of
beam diffraction and self-imaging effects are clearly observed when
period and amplitude of forcing satisfy the condition $\Gamma \equiv
F_0 a / (\lambdabar \omega)=2.405$ (first zero of Bessel $J_0$
function), as shown in Figs.2(e) and (f).

\section{Wave packet dynamics at the $\mathcal{PT}$ symmetry-breaking point}

 As $\alpha$ is increased to reach and cross the $\mathcal{PT}$ symmetry-breaking point
$\alpha_c$, gap narrowing till band merging, associated to the
appearance of pairs of complex-conjugate eigenvalues, is observed
\cite{Makris08}. For instance, for the potential defined by Eq.(2),
at the transition point the band diagram is given by the
free-particle energy dispersion curve $E=\lambdabar^2
\kappa^2/(2n_s)$, periodically folded inside the first Brillouin
zone \cite{Makris08} (see the dotted curve in Fig.1). In this case,
as previously noticed for BOs \cite{Longhi09}, wave packet transport
is deeply modified and can not be described by means of the
canonical model (3) introduced in the previous section. From a
physical viewpoint, this is related to the highly non-reciprocal
behavior of Bragg scattering in the crystal and violation of
Friedel's law of Bragg diffraction for crystal inversion
\cite{Keller97,Berry98}. As in Ref. \cite{Longhi09}, we limit here
to consider the dynamical behavior of a broad wave packet in the
complex lattice defined by Eq.(2) at $\alpha=\alpha_c=1$, i.e.
$V(x)=V_0 \exp(i k_B x)$, which enables a rather simple analytical
and physical analysis. As opposed to Ref. \cite{Longhi09}, we assume
here an ac-like force, namely $F(z)=F_0 \cos(\omega z)$. Figure 3
shows a few typical examples of wave packet dynamics as obtained by
a numerical analysis of Eq.(1) when the lattice is excited at $z=0$
with a broad Gaussian beam $\psi(x,0)=\exp(-x^2/w^2)$  for a fixed
value of the ac modulation period $\Lambda= 2 \pi/ \omega$ and for
increasing values of the amplitude $F_0$. For a small amplitude
$F_0$, it turns out that the wave packet propagates as if the
lattice were absent [Fig.3(a)], following an oscillatory path as
predicted by the semiclassical analysis of Eq.(1) with $V=0$. Owing
to the external force, the mean wave packet momentum varies
periodically according to $\lambdabar k(z)=\int_0^z d \xi
F(\xi)=(F_0/\omega)\sin(\omega z)$ . As the forcing $F_0$ reaches
and crosses the critical value $F_c=k_B \lambdabar \omega/2$, new
wave packets, arising from first-order Bragg diffraction,
periodically bifurcate from the primary beam at propagation
distances $z$ satisfying the Bragg condition $k(z)=-k_B/2$ [see
Figs.3(b) and 3(c)]. The mean momentum of the first-order diffracted
wave packets differs from that of the primary wave packet by an
additional term $k_B$, which explains the refraction of the
first-order diffracted beams at the angle $\theta=dx/dz=k_B
\lambdabar/n_s$ observed in Figs.3(b) and 3(c). At stronger forcing,
namely for $F \geq 3F_c$, additional wave packets bifurcate from the
first-order wave packets at propagation distances $z$ such that
$k(z)=-3k_B/2$, as shown in Fig.3(d). These wave packets originate
from second-order Bragg diffraction in the crystal and greatly
complicate the pattern scenario. The mean momentum of second-order
diffracted wave packets differs from that of the primary wave packet
by the additional term $2k_B$, which explains the larger (twice)
refraction angle of second-order diffracted beams as compared to
that of first-order diffracted beams [see Fig.3(d)]. At even higher
forcing, i.e. at $F \geq 5F_c$, new wave packets bifurcate from the
second-order wave packets because of third-order Bragg diffraction
at propagation distances $z$ such that $k(z)=-5k_B/2$, and so on.
The appearance of  wave packets generated by Bragg diffraction at
various orders is rather abrupt, as indicated by the behavior of the
normalized beam power $P(z)=\int dx |\psi(x,z)|^2 /
\int dx |\psi(x,0)|^2$ versus propagation distance $z$ shown in the right panels of Fig.3.\\
The dynamical scenario observed in numerical simulations can be
analytically captured by considering the limiting case of a plane
wave exciting the crystal at $z=0$ with initial wave number $k=0$.
In fact, the solution to Eq.(1) with the initial condition
$\psi(x,0)=1$ is given by the superposition of diffracted plane
waves at different Bragg orders according to
\begin{equation}
\psi(x,z)=\sum_{n=0}^{\infty} a_n(z) \exp [ik(z)x+ink_Bx-i
\gamma_n(z)].
\end{equation}
In Eq.(7) we have set
\begin{eqnarray}
k(z) & = & \frac{1}{  \lambdabar} \int_0^z d \xi F(\xi)=\frac{F_0}{
\lambdabar \omega} \sin(\omega z) \\
\gamma_n(z) & = & \frac{\lambdabar}{2 n_s} \int_0^z d \xi \left
[nk_0+k(\xi) \right]^2,
\end{eqnarray}
whereas the amplitudes $a_n(z)$ are calculated from the recurrence
relations
\begin{equation}
a_{n}(z)=-i \frac{V_0}{\lambdabar}\int_0^z d \xi a_{n-1}(\xi) \exp
\left[ i \gamma_{n}(z)-i\gamma_{n-1}(z) \right]
\end{equation}
with $a_0(z)=1$. The amplitude $a_0$ corresponds, in Fig.3, to the
primary wave packet, and the independence of $a_0$ from $z$
indicates that this wave packet propagates as if the lattice were
absent. The amplitude $a_1$ corresponds to the first-order
diffracted wave packets, $a_2$ to second-order diffracted wave
packets, and so on. According to Eq.(10), a wave packet
corresponding to Bragg diffraction at order $n$ bifurcates from a
wave packet of order $(n-1)$, and Bragg diffraction is effective
provided that the phase difference
$\varphi_n(z)=\gamma_{n}(z)-\gamma_{n-1}(z)$ entering in the
exponential of the integral on the right-hand-side of Eq.(10) has a
stationary point. The condition $d  \varphi_n(z) /dz=0$ is satisfied
at propagation distances $z_0$ such that
\begin{equation}
k(z_0)=-k_B \left(n-\frac{1}{2} \right) \; \; (n=1,2,3,...).
\end{equation}
\begin{figure}
  \includegraphics[width=85mm]{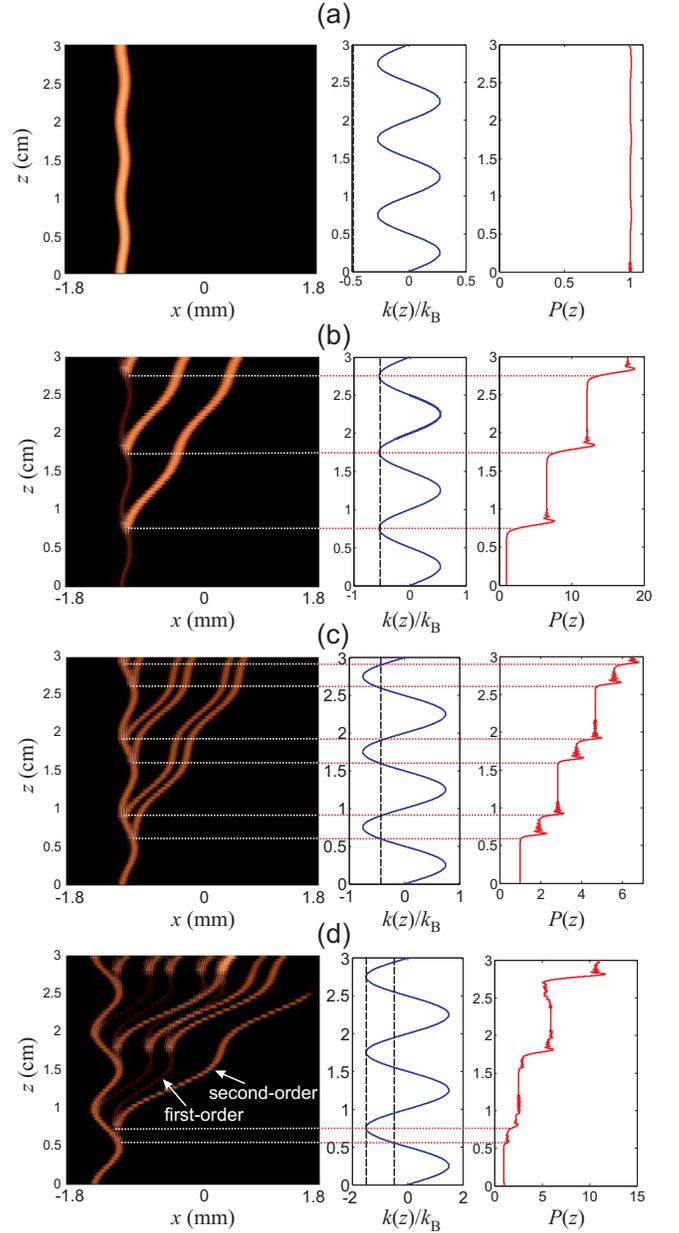}\\
   \caption{(color online) Propagation of a broad Gaussian wave packet (input beam size $w=80 \;\mu$m) in the complex lattice $V(x)=V_0 \exp(2 \pi i x/a)$
   for $V_0=0.0002$, $a=6 \; \mu$m, $\lambda=633$ nm, $n_s=1.42$, subjected to a sinusoidal ac-like force $F(z)=F_0 \cos (2 \pi z / \Lambda)$ with period
   $\Lambda=1 \; \mathrm{cm}$ and with increasing values of force amplitude
   $F_0$: (a) $F_0/F_c=0.5$, (b) $F_0/F_c=1$, (c) $F_0/F_c=1.5$, and (d)
   $F_0/F_c=3$, where the critical forcing $F_c$ is defined by
   Eq.(15). Left panels show snapshots of $|\psi(x,z)|$; the central panels depict the
   behavior of $k(z)$, normalized to the Bragg wave number $k_B$, as given by Eq.(8); the right panels show the
   evolution of normalized beam power $P(z)$. The horizontal dotted lines indicate the
   crossing points $z=z_0$ where new wave packets are generated by Bragg scattering. For the sake of clarity, in (d)
   only the first two crossing points are indicated. At the crossing points a
   rather abrupt increase of beam power $P(z)$ is observed. The crossing is linear in (b) and (d), and parabolic in (c).}
\end{figure}
i.e. for a wave number that reaches the edge of the Brillouin zone.
Note, however, that in Eqs.(7) and (11) $n$ is a positive (but not a
negative) integer number: this circumstance is a clear signature
that Bragg scattering is highly non-reciprocal, a feature which is
peculiar to the complex nature of the crystal
\cite{Keller97,Berry98}.
 If the crossings of the
stationary points are fast enough, $a_n(z)$ is basically constant
far from the stationary points, with abrupt changes at $z=z_0$,
namely
\begin{equation}
a_n(z_0^+) \simeq  a_n(z_0^-) + R a_{n-1} (z_0)
\end{equation}
 where
\begin{equation}
R = -i\frac{V_0}{\lambdabar}\int d \xi \exp [i \varphi_n(\xi)]
\end{equation}
and the integral on the right hand side of Eq.(13) is extended to
the neighborhood of $z_0$. From these results one can readily
explain the abrupt changes of the total beam power $P(z)$ observed
in Fig.3 (right panels) at the crossing points shown in the central
panels of Fig.3, the existence of a critical forcing $F_c$ below
which no bifurcating wave packets appear, as well as the occurrence
of higher-order bifurcating wave packets and a dynamical scenario
with increasing complexity as the amplitude of forcing is increased.
In fact, according to Eqs. (8) and (11) the stationary points which
generate the bifurcating wave packets of order $n$ are obtained from
the equation
\begin{equation}
\frac{ F_0}{ \lambdabar \omega} \sin(\omega
z_0)=-\frac{k_B}{2}(2n-1),
\end{equation}
which can be satisfied provided that the amplitude $F_0$ of forcing
is larger than $F_c(2n-1)$, where
\begin{equation}
F_c= \frac{1}{2 }\lambdabar \omega k_B
\end{equation}
is the critical forcing amplitude. Therefore, for $F_0 < F_c$ no
diffracted wave packets of any order are generated [Fig.3(a)], for
$F_c \leq F_0<3F_c$ first-order diffracted wave packets are
generated, for $3 F_c \leq F_0<5F_c$ first-order and second-order
wave packets are generated, and so on. The graphical determination
of the stationary points is depicted in the central panels of Fig.3
as the crossing of the sinusoidal curve $k(z)/k_B$ with the vertical
dashed lines $k/k_B=-0.5$ (for first-order Bragg diffraction) and
$k/k_B=-1.5$ [for second-order Bragg diffraction, shown solely in
Fig.3(d)]. Let ut discuss in some detail the generation of
first-order diffracted wave packets ($n=1$). For $F_c<F_0<3F_c$
[Fig.3(c)], in each ac oscillation cycle there are two stationary
points, and the crossing is linear. Near each of the stationary
points, $ \varphi_1(z)$ can be thus approximated as
$\varphi_1(z)=\varphi_1(z_0)+(1/2) (d^2\varphi_1/dz^2) (z-z_0)^2$,
where the derivative $(d^2\varphi_1/dz^2)$ is calculated at the
stationary point $z=z_0$. From Eq.(13), it follows that the
amplitude factor $|R|$ of the generated diffracted beam after each
stationary phase point can be approximately computed as
\begin{eqnarray}
|R| & \simeq & \frac{V_0}{\lambdabar} \left|
\int_{-\infty}^{\infty}d \xi \exp \left( \frac{i}{2} \frac{d^2
\varphi_1}{ d z^2} \xi^2 \right) \right| \nonumber
\\
& = & \frac{V_0}{\lambdabar} \sqrt{\frac{ 2 \pi n_s}{k_B
\sqrt{F_0^2-F_c^2}}}.
\end{eqnarray}
 In deriving
Eq.(16), we have taken into account that $(d^2 \varphi_1
/dz^2)=F(z_0)k_B/n_s$, $|F(z_0)|=(F_0^2-F_c^2)^{1/2}$  and
$\int_{-\infty}^{\infty}d \xi \exp(i \xi^2)= \sqrt{i \pi}$. For
parameter values used in the simulations of Fig.3(c), from Eq.(16)
one has $|R| \simeq 0.95$, which is in excellent agreement with the
staircase behavior of beam power $P(z)$ shown in the right panel of
Fig.3(c). Equation (16) fails to predict the correct amplitude
factor $R$ when $F_0 \rightarrow F_c^{+}$, i.e. when the two
crossing points in each ac oscillation cycle coalesce. For
$F_0=F_c$, the crossing is parabolic [see Fig.3(b), central panel],
and $ \varphi_1(z)$ in Eq.(13) should now be approximated as
$\varphi_1(z)=\varphi_1(z_0)+(1/6) (d^3\varphi_1/dz^3) (z-z_0)^3$,
where the derivative $(d^3\varphi_1/dz^3)$ is calculated at the
crossing point $z=z_0$. In this case, in place of Eq.(16) one has
\begin{eqnarray}
|R| & \simeq & \frac{V_0}{\lambdabar} \left|
\int_{-\infty}^{\infty}d \xi \exp \left( \frac{i}{6} \frac{d^3
\varphi_1}{ d z^3} \xi^3 \right) \right| \nonumber
\\
& = & \frac{V_0}{\lambdabar} 2 \pi \mathrm{Ai}(0) \left( \frac{4
n_s}{\lambdabar \omega^2 k_B^2}\right)^{1/3}
\end{eqnarray}
where $\mathrm{Ai}(\xi)$ is the Airy function. In deriving Eq.(17),
we have taken into account that $(d^3 \varphi_1 /dz^3)=(k_B/n_s)
(dF/dz)=\lambdabar k_B^2 \omega^2 /(2n_s)$ at $z=z_0$, and
$\int_{-\infty}^{\infty}d \xi \exp(i \xi^3/3)= 2 \pi
\mathrm{Ai}(0)$. For parameter values used in the simulations of
Fig.3(b), from Eq.(17) one has $|R| \simeq 2.245$. Note that, with
this amplitude factor, the staircase behavior of beam power $P(z)$
shown in the right panel of Fig.3(b) is reproduced with excellent
accuracy.

\section{Conclusions}
The coherent motion of a Bloch wave packet in a tight-binding
lattice driven by an ac electric field is known to show a
self-imaging effect that arises from quasi-energy band collapse of
the time-periodic Hamiltonian. This phenomenon, referred to as
dynamic localization \cite{Dunlap86,Holthaus92}, has been recently
observed for both matter and optical waves as a suppression of wave
packet broadening in the lattice. In this work, we investigated
theoretically the behavior of a Bloch particle in a complex crystal
with $\mathcal{PT}$ symmetry subjected to a sinusoidal ac-like
force. As compared to ordinary crystals, complex crystals exhibit
some unique properties, such as violation of the Friedel's law of
Bragg scattering and nonreciprocal diffraction. For an unbroken
$\mathcal{PT}$ symmetry and in the single-band approximation, it has
been shown that the quasi-energy spectrum of the time-periodic
non-Hermitian Hamiltonian remains real-valued. In this regime, like
in an ordinary crystal exact band collapse is possible within the
NNTB approximation, i.e. for a sinusoidal band shape. At the
$\mathcal{PT}$ symmetry breaking transition point, band merging
greatly modifies the wave packet dynamics as compared to an ordinary
crystal. Here we have investigated in details the dynamics of a
broad wave packet in the $\mathcal{PT}$-symmetric potential
$V(x)=V_0 \exp(2 \pi i x/a)$ and shown that the complexity of the
dynamical scenario greatly increases as the strength of forcing is
increased, with the appearance of a cascading of bifurcating wave
packets. The main features observed in numerical simulations have
been analytically explained by considering the Bragg scattering of a
plane wave in the complex lattice subjected to the sinusoidal ac
driving force. These results are complementary to the recent study
of Bloch oscillations in complex crystals \cite{Longhi09}, and are
expected to motivate further investigations aimed to explore the
exotic transport properties o complex crystals.


\end{document}